# The universality of the Carnot theorem

**Julian Gonzalez-Ayala and F. Angulo-Brown**

Departamento de Física, Escuela Superior de Fisica y Matematicas, Instituto Politecnico Nacional, Edif. No 9, U.P. Zacatenco, 07738 Mexico, DF, Mexico

E-mail: noldor_21@yahoo.com.mx and angulo@esfm.ipn.mx

**Abstract.** It is common in many thermodynamic textbooks to illustrate the Carnot theorem through the usage of diverse state equations for gases, paramagnets, and other simple thermodynamic systems. As it is well-known, the universality of the Carnot efficiency is easily demonstrated in a temperature-entropy diagram, which means that $\eta_C$ is independent of the working substance. In this present work we remark that the universality of the Carnot theorem goes beyond the conventional state equations, and it is fulfilled by gas state equations that do not behave as ideal gas in the dilution limit namely $V \to \infty$. Some of these unconventional state equations have certain thermodynamic "anomalies" that nonetheless do not forbid them from obeying the Carnot theorem. We discuss how this very general behaviour arises from the Maxwell relations, which are connected with a geometrical property expressed through preserving area transformations. A rule is proposed to calculate the Maxwell relations associated with a thermodynamic system by using the preserving area relationships. In this way it is possible to calculate the number of possible preserving area mappings by giving the number of possible Jacobian identities between all pairs of thermodynamic variables included in the corresponding Gibbs equation. This work is intended for undergraduate and specialists in thermodynamics and related areas.

**1. Introduction**

As it is well known, if the Carnot theorem is demonstrated in a temperature-entropy $(T - S)$ plane, one obtains as a corollary, that the Carnot efficiency given by $\eta_C = 1 - T_2/T_1$, is independent of the working substance, where $T_2$ is the absolute temperature of a cold reservoir and $T_1$ the corresponding temperature of a hot reservoir. However, in many thermodynamics textbooks [1-4] and also in some articles [5, 6] this efficiency is calculated for several working substances by means of their corresponding state equations in the work plane $(Y - X)$, where $Y$ is an intensive thermodynamic variable and $X$ the corresponding extensive conjugate variable. For gases, $Y$ and $X$ are the pressure $p$ and the volume $V$ while for paramagnets they are the magnetic field intensity $H$ and the magnetization $M$, respectively. Many other simple thermodynamic systems have been subjected to this procedure (elastic wire, surface film, etc.). Obviously, an individual demonstration substance by substance in the work plane is not necessary, however, many authors judge illustrative to make this kind of exercise. In all known cases, the working substances are familiar physical systems. Recently, Penrose asserted [7] that: "this law (the second law of thermodynamics) has a universality that goes beyond any particular system of dynamical rules that one might be concerned with… It applies also to hypothetical dynamical theories that we have no good reason to believe they have relevance to the actual universe that we inhabit…" That is, the universality of the Carnot theorem (taken as an expression of the second law) goes far beyond any particular system. This problem concerning the applicability of macroscopic laws of thermodynamics also arise in other contexts, such as the behaviour of isothermal compressibility, $\kappa_T$, in the heterogeneous region in a Van der Waals isotherm (the $S$-shaped region). As is well known, part of this region has a positive slope of the curve in a $p - V$ plane, leading to $\kappa_T < 0$. Though, from experience we know that $\kappa_T > 0$ for most substances; this fact is not implied by either the first law or the second law [8]. Thus, it is possible to find thermodynamic systems with certain behaviours not very common in nature, but that are not forbidden by the thermodynamics laws. As a matter of fact, the universality extent of the second law is impressive. In the present article we calculate the Carnot efficiency for several "state equations" that do not represent usual substances. This is done by means of a state equation for gases which includes the virial expansion and also some "uncommon" state equations. In 1973, Tykodi and Hummel (T-H) [9] proposed a simple procedure to distinguish between state equation of gases that are consistent with the first and second laws and those which are not. Among the uncommon equations generated by the mentioned equation, we find some that do not fulfil the T-H criterion of thermodynamic consistency. Nevertheless, they obey the Carnot theorem. In this work we also discuss other strange properties of the so-called uncommon equations.

In 2001, Ambegaokar and Mermin [10] pointed out that the way to look at the Maxwell relations within the context of the Jacobian identities is little known in thermodynamic standard textbooks. We believe that in general this is the case, with a few exceptions like [11]. The present paper discusses geometrical properties of thermal loops based on Jacobian identities and Legendre transformations. From a formal point of view, the Maxwell relations, as it will be discussed later are somehow responsible of the known expression for the Carnot efficiency and they summarize the first and second laws of thermodynamics for reversible closed processes. All the uncommon state equations presented fulfil $J(p, V/T, S) = 1$, just as in common state equations. This work has been developed in a self-contained manner in order to be accessible for undergraduate students because several concepts treated here are helpful to understand the first and second laws of thermodynamics. They are also helpful for specialists in thermodynamics and related areas.

The present article is organized as follows: Section 2 introduces a somewhat general state equation for gases, and we demonstrate that it fulfils the Carnot theorem. Section 3 summarizes the T-H criterion of thermodynamic consistency and we apply it to several unconventional state equations; in particular, the behaviour of one of them is analysed. Section 4 discusses a preserving area property related to Maxwell relations and also in connection with the Carnot efficiency. Finally, some conclusions are presented.

**2. A somewhat general state equation for gases**
The so-called virial state equation is given by

$$p = RT\left(\frac{1}{v} + \frac{B_2}{v^2} + \frac{B_3}{v^3} + \ldots\right), \tag{1}$$

where $R$ is the universal gas constant, $v$ is the molar volume ($v = V/n$, $n$ is the molar number) and $B_2$, $B_3$, etc. are the second, third, etc. virial coefficients and are functions of the temperature only. This equation encompasses every gas state equation that has the property to converge to the ideal gas state equation in the limit as $V \to \infty$; that is, for diluted gases. For example, equation (1) reproduces the ideal gas equation when

$B_2 = B_3 = \ldots = 0$; and reproduces the Van der Waals equation for $B_2 = b - a/RT$ and $B_3 = B_4 = \ldots = 0$, where $a$ and $b$ are the usual Van der Waals parameters. In 1990, Agrawal and Menon showed how the Van der Waals equation verifies the Carnot efficiency [5]. Later, in 2006 Tjiang and Sutanto [6] made the same verification for the Redlich-Kwong state equation. In the same article, the authors proposed a method to calculate the Carnot efficiency for an arbitrary state equation for gases. This is true for well-behaved state equations; although this could not be the case for unconventional or even unrealistic substances. For example, those whose limit for diluted gases is different from the ideal gas state equation, such as the following equation,

$$p = \frac{aV}{T}, \qquad (2)$$

where $a$ is a positive constant with proper units. Hereafter $a$ is not the Van der Waals parameter. At first glance this is a non-physical state equation for a gas, because when T grows p diminishes and when the gas expands the pressure increases. Also, in the limit of diluted gas it does not reproduce the behaviour of the ideal gas; nor is the energy well-behaved as we shall see below. Additionally, it can be said that the virial state equation cannot reproduce equation (2). In order to cover a wider set of equations than the virial expansion we will work with the following equation,

$$p = \sum_{j=-n}^{n} \sum_{i=-m}^{m} a_{ij}(T - T_0)^i (V - V_0)^j, \qquad (3)$$

where $m$ and $n$ are positive integers ($V_0, T_0$ is a reference state) and $a_{ij}$ are the constant coefficients of the series (perhaps the only condition that we would impose on the last equation is that $i$ must be different from zero, otherwise there is no relationship between the variables $p$ and $V$ with $T$ and $S$). In fact, any particular equation derived from this equation also fulfils the Carnot theorem as will be shown. Equation (3) with $T_0 = V_0 = 0$, $a_{-1,1} = 1$ and any other $a_{ij} = 0$ gives equation (2). In the same way, many equations can be constructed, as for example, when $T_0 = V_0 = 0$, $a_{1,-1} = a$ and $a_{2,-3} = b$ (hereafter $b$ is not the Van der Waals parameter referred above), and every other $a_{ij} = 0$ results in

$$p = \frac{aT}{V} + \frac{bT^2}{V^3}. \qquad (4)$$

Similarly, with $T_0 = V_0 = 0$, $a_{2,-2} = a$ and $a_{4,-5} = b$ and every other $a_{ij} = 0$ we obtain

$$p = \frac{aT^2}{V^2} + \frac{bT^4}{V^5}, \qquad (5)$$

and with $T_0 = V_0 = 0$, $a_{2,0} = a$ and $a_{0,-2} = b$ and every other $a_{ij} = 0$ then

$$p = aT^2 - \frac{b}{V^2}, \qquad (6)$$

in all these cases $a$ and $b$ are constants with proper units. It should be said that Eq. (3) covers a wide set of possible functions, for example, considering an analytical function $\mathcal{F}(T,V)$, thus this function can be expressed by a power series such as the following one,

$$\mathcal{F}(T,V) = \sum_{i=0}^{\infty} \sum_{j=0}^{\infty} b_{ij}(T - T_0)^i (V - V_0)^j,$$

which can be represented by Eq. (3) when $n, m \to \infty$ and every $a_{ij} = 0$ for $i, j < 0$.

The virial expansion is a power series of the inverse of the volume $V$, where the coefficients of the expansion are functions of the temperature. As far as we know [12, 13] those coefficients can be written also in a power series of the temperature $T$, in this way we can write the virial expansion as (see Appendix A),

$$p = \sum_{j=0}^{\infty} \sum_{i=1}^{\infty} a_{ij} T^i V^{-j}, \tag{A10}$$

which is a particular case of equation (3). In fact, equation (3) reproduces many gas state equations which give the ideal gas behaviour when $V \to \infty$ (diluted gas), among other forms that do not reproduce the ideal gas limit when $V \to \infty$.

In addition, equation (3) fulfils the Carnot theorem. The following is a well-known thermodynamical identity [2],

$$\left(\frac{\partial U}{\partial V}\right)_T = T\left(\frac{\partial p}{\partial T}\right)_V - p, \tag{7}$$

where $U$ is the total internal energy. Since $dU$ is exact then

$$\left(\frac{\partial^2 U}{\partial V \, \partial T}\right) = T\left(\frac{\partial^2 p}{\partial T^2}\right)_V.$$

Integration over the variables V and T yields to

$$U = \iint \left(\frac{\partial^2 U}{\partial V \, \partial T}\right) dV dT + \int f(T) \, dT + H(V), \tag{8}$$

where f(T) and H(V) are functions to be determined. At this point we can impose the condition that a very dilute monoatomic gas behaves as the ideal gas, that is,

$$\lim_{V \to \infty} U = \frac{3}{2} nRT,$$

where n is the molar gas number, then,

$$f(T) = \frac{3}{2} nR.$$

In this way, by using equations (3), (7) and (8) we arrive at the following result,

$$U = \sum_{\substack{i=-m \\ i \neq 0}}^{m} \left((i-1)T - T_0\right)(T - T_0)^{i-1} \left(\sum_{\substack{j=-n \\ j \neq -1}}^{n} a_{ij} \frac{(V - V_0)^{j+1}}{j+1} + a_{i,-1} \ln(V - V_0)\right) + F(T) + U_0, \tag{9}$$

where F(T) is an arbitrary temperature function only (for the ideal gas $F(T) = 3nRT/2$).

Now we calculate the efficiency of a Carnot cycle in a $(p, V)$ diagram for equation (3). The graphic aspect of this cycle is undetermined, but figures 2(c), 3(c), 4(c) and 5(c) are particular cases arising from equation (3).

The efficiency calculation requires knowing the heats involved in the cycle, since the relationship between the efficiency and the heats is

$$\eta = |W|/Q_{in} = 1 + Q_{out}/Q_{in}, \qquad (10)$$

where $W$ is the total work, $Q_{in}$ is the positive heat input and $Q_{out}$ the negative heat output. Here we use the convention that $Q_{out}$ is negative when leaving the system and similarly for the total work $W$ (negative when the system makes work and positive when the work is made over the system). In an isothermal process the inexact differential of heat $dQ$ can always be written as [2],

$$dQ = T\left(\frac{\partial p}{\partial T}\right)_V dV, \qquad (11)$$

by introducing equation (3) in (11) and integrating over the volume we calculate the heat $Q$ for the proposed state equation (3),

$$Q = \sum_{i=-m}^{m} iT(T-T_0)^{i-1}\left(\sum_{\substack{j=-n \\ j\neq -1}}^{n} a_{ij}\frac{(V-V_0)^{j+1}}{j+1} + a_{i,-1}\ln(V-V_0)\right). \qquad (12)$$

Then equation (10) is,

$$\eta = 1 + \frac{\sum_{i=-m}^{m} iT_2(T_2-T_0)^{i-1}\left(\sum_{\substack{j=-n \\ j\neq -1}}^{n} a_{ij}\frac{(V_4-V_0)^{j+1}-(V_3-V_0)^{j+1}}{j+1} + a_{i,-1}\ln\left(\frac{V_4-V_0}{V_3-V_0}\right)\right)}{\sum_{i=-m}^{m} iT_1(T_1-T_0)^{i-1}\left(\sum_{\substack{j=-n \\ j\neq -1}}^{n} a_{ij}\frac{(V_2-V_0)^{j+1}-(V_1-V_0)^{j+1}}{j+1} + a_{i,-1}\ln\left(\frac{V_2-V_0}{V_1-V_0}\right)\right)}, \qquad (13)$$

where we have used the labels shown in figure 1 only to illustrate the processes order.

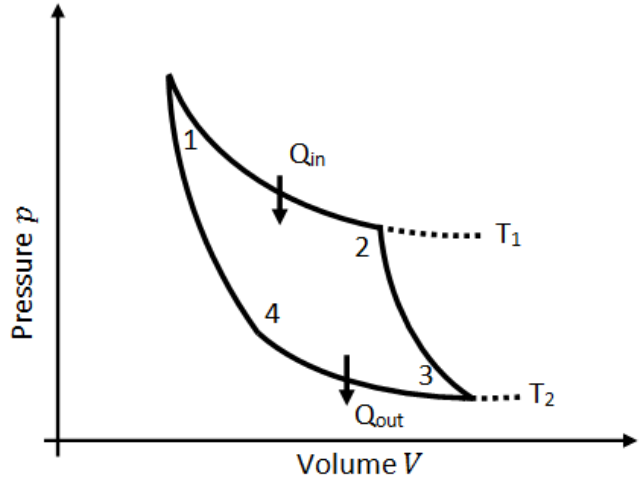

**Figure 1.** Carnot cycle in the work plane ( $(p,V)$ plane) for the ideal gas case.

Now we consider the inexact differential of heat for an adiabatic process [2],

$$\left(\frac{\partial U}{\partial T}\right)_V dT + T\left(\frac{\partial p}{\partial T}\right)_V dV = 0, \qquad (14)$$

which has the integrating factor $\mu = \frac{1}{T}$ [6] that transforms the inexact differential into an exact one. By using equations (3) and (9), the solution for equation (14) multiplied by the integrating factor $\mu$ is,

$$\sum_{i=-m}^{m} i(T-T_0)^{i-1} \left( \sum_{\substack{j=-n \\ j \neq -1}}^{n} a_{ij} \frac{(V-V_0)^{j+1}}{j+1} + a_{i,-1} \ln(V-V_0) \right) + J(T) = \text{constant}, \tag{15}$$

where $J(T) = \int \frac{F'(T)}{T} dT$ is a temperature function only. Over the adiabatic process $2 \to 3$ the constant of equation (15) is the same because they are on the same adiabatic curve, then,

$$J(T_2) - J(T_1) + \sum_{i=-m}^{m} i(T_2 - T_0)^{i-1} \left( \sum_{\substack{j=-n \\ j \neq -1}}^{n} a_{ij} \frac{(V_3 - V_0)^{j+1}}{j+1} + a_{i,-1} \ln(V_3 - V_0) \right)$$

$$- \sum_{i=-m}^{m} i(T_1 - T_0)^{i-1} \left( \sum_{\substack{j=-n \\ j \neq -1}}^{n} a_{ij} \frac{(V_2 - V_0)^{j+1}}{j+1} + a_{i,-1} \ln(V_2 - V_0) \right) = 0, \tag{16}$$

for the adiabatic process $4 \to 1$ the constant in equation (15) is also the same, then,

$$J(T_1) - J(T_2) + \sum_{i=-m}^{m} i(T_1 - T_0)^{i-1} \left( \sum_{\substack{j=-n \\ j \neq -1}}^{n} a_{ij} \frac{(V_1 - V_0)^{j+1}}{j+1} + a_{i,-1} \ln(V_1 - V_0) \right)$$

$$- \sum_{i=-m}^{m} i(T_2 - T_0)^{i-1} \left( \sum_{\substack{j=-n \\ j \neq -1}}^{n} a_{ij} \frac{(V_4 - V_0)^{j+1}}{j+1} + a_{i,-1} \ln(V_4 - V_0) \right) = 0. \tag{17}$$

From equations (16) and (17) after some algebra the following result is found,

$$\sum_{i=-m}^{m} i(T_2 - T_0)^{i-1} \left( \sum_{\substack{j=-n \\ j \neq -1}}^{n} a_{ij} \frac{(V_4 - V_0)^{j+1} - (V_3 - V_0)^{j+1}}{j+1} + a_{i,-1} \ln\left(\frac{V_4 - V_0}{V_3 - V_0}\right) \right) =$$

$$- \sum_{i=-m}^{m} i(T_1 - T_0)^{i-1} \left( \sum_{\substack{j=-n \\ j \neq -1}}^{n} a_{ij} \frac{(V_2 - V_0)^{j+1} - (V_1 - V_0)^{j+1}}{j+1} + a_{i,-1} \ln\left(\frac{V_2 - V_0}{V_1 - V_0}\right) \right).$$

Finally, by factorizing $\frac{T_2}{T_1}$ in equation (13) we arrive at the desired result,

$$\eta = 1 - \frac{T_2}{T_1}. \tag{18}$$

Equation (15) represents the entropy of the system, so that in an indirect way the previous calculation is equivalent to the Carnot efficiency demonstration made in the $(T, S)$ plane. Nevertheless, it is made by using explicitly the state equation and it rather represents a completeness exercise for the purpose of this work.

Thus, we now properly write the entropy for equation (3),

$$S = \sum_{i=-m}^{m} i(T - T_0)^{i-1} \left( \sum_{\substack{j=-n \\ j \neq -1}}^{n} a_{ij} \frac{(V - V_0)^{j+1}}{j + 1} + a_{i,-1} \ln(V - V_0) \right) + \int \frac{F'(T)}{T} dT + S_0, \tag{19}$$

where F(T) is an arbitrary temperature function only (for the ideal monoatomic gas $F(T) = 3nRT/2$). That explains the simple nature of the integrating factor $\mu$ that transforms the inexact differential of heat into the exact differential of entropy.

Equations (2), (4) and (6) fulfil the correct form of the Carnot efficiency because equation (3) fulfils it also. In particular, as $V \to \infty$ equation (6) reduces to $p = aT^2$; that is, we have an equation that in the diluted limit does not reproduce the ideal gas equation, nevertheless it satisfies the Carnot theorem. As we will see in the next section, some of these equations do not fulfil the Tykodi-Hummel criterion for thermodynamic consistency mentioned in the Introduction.

## 3. On the Tykodi-Hummel criterion

In their work Tjiang and Sutanto showed that an arbitrary gas equation of state fulfils the Carnot theorem. But, how arbitrary can the equation of state be? In the last section we consider a somewhat general equation of state and regardless the not very familiar contributions of positive powers of the volume $V$ and negative powers of the temperature $T$, it does not exclude them from being thermodynamically well behaved functions. It is not until the invalidity of those equations is established (within the frame of the T-H criterion) that the Carnot theorem may be tested for those functions.

State equations for several thermodynamic systems (as fluids, paramagnets, etc.) are built following empirical and semi-empirical procedures, or methods based on first principles derived from statistical mechanics, as is the case of the virial equation. Despite this, there are not many works on explaining whether the existing state equations are compatible with the first and the second laws of thermodynamics (because it is not expected that these equations describe the behaviour of gases close to zero temperature, in the T-H criterion compatibility with third law is not considered).

On this subject Tykodi and Hummel [9] proposed a way to verify the thermodynamic validity of any gas state equation. The result obtained was that if any state equation of the form $p = p(T, V)$ is expected to be in agreement with the first and the second laws of thermodynamics, where p is the pressure, T the temperature and V the volume, the state equation should be expressed as,

$$[p + \xi(T, V)]\phi(V) = RT, \tag{20}$$

where $\phi(V)$ is a volume function, $R$ the universal gas constant and $\xi(T, V)$ is a temperature and volume function defined as follows,

$$\xi(T, V) = -T \int T^{-2} \left( \frac{\partial U}{\partial V} \right)_T dT, \tag{21}$$

where the integral is performed at constant V. Equations (20) and (21) derive from well-known simple thermodynamic identities. From the Gibbs equation,

$$dU = TdS - pdV,$$

it immediately follows that,

$$\left(\frac{\partial U}{\partial V}\right)_T = T\left(\frac{\partial S}{\partial V}\right)_T - p,$$

by using the following Maxwell relation $\left(\frac{\partial S}{\partial V}\right)_T = \left(\frac{\partial p}{\partial T}\right)_V$ the above equation can be written as

$$\left(\frac{\partial U}{\partial V}\right)_T = T\left(\frac{\partial p}{\partial T}\right)_V - p,$$

that can be rearranged as,

$$\left(\frac{\partial U}{\partial V}\right)_T = T^2\left(\frac{\partial \frac{p}{T}}{\partial T}\right)_V. \tag{22}$$

Integration of equation (22) at constant temperature gives Eqs (20) and (21) [9]. In addition, it is expected that in the limit of a very dilute gas, it behaves as an ideal gas, that is,

$$\lim_{V\to\infty} p = p_{\text{ideal gas}}, \tag{23}$$

in other words,

$$\lim_{V\to\infty}\frac{V}{\phi(V)} = 1, \tag{24}$$
$$\lim_{V\to\infty} V\xi(T,V) = 0.$$

Equation (21) can be rewritten as,

$$\xi(T,V) = -T\int T^{-2}\left(T\left(\frac{\partial p}{\partial T}\right)_V - p\right)dT. \tag{25}$$

In this way, all we need is to know the state equation to evaluate equation (25) and additionally to obtain the function $\phi(V)$ to determine whether such state equation is thermodynamically consistent or not. The T-H criterion has been applied to confirm the thermodynamic consistency of gas state equations like the ideal gas (where $\phi(V) = V$ and $\xi(T,V) = 0$), Van der Waals (where $\phi(V) = v - b$ and $\xi(T,V) = a/v^2$), Redlich-Kwong (where $\phi(V) = v - b$ and $\xi(T,V) = a/(v(v-b)T^{1/2})$) and a few more [9].

In Figure 2 we depict a $p - V$ diagram of adiabatic processes (figure 2a), isothermal processes (figure 2b) and a Carnot cycle (figure 2c) for a substance described by equation (2).

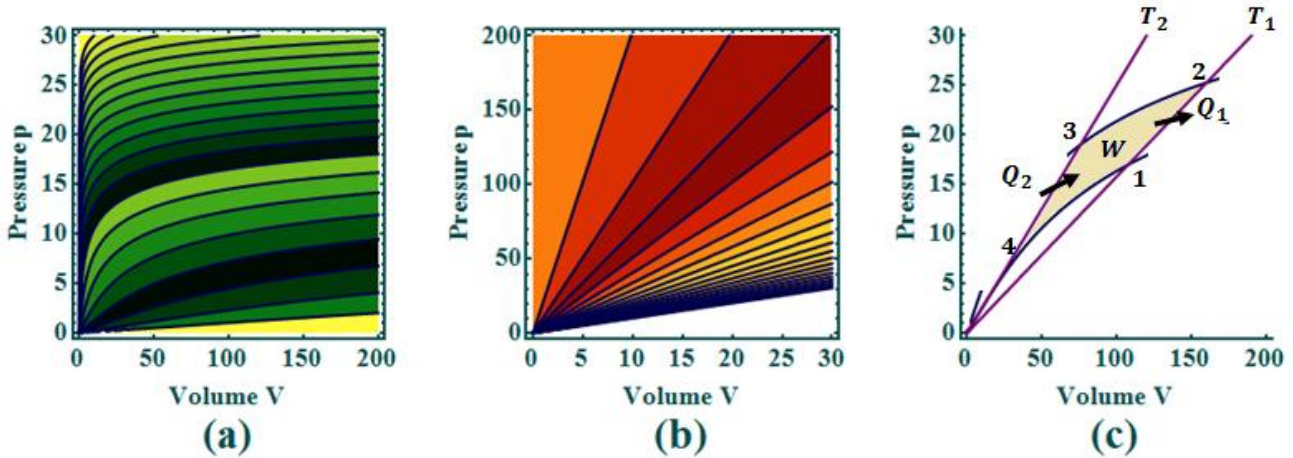

**Figure 2.** (a) Adiabatic processes, (b) isothermal processes and (c) a possible Carnot cycle for the state equation $p = \frac{aV}{T}$.

Obviously, we can construct Carnot cycles for equations (4), (5) and (6) (see figures (2), (3) and (4), respectively). In particular, equations (2), (5) and (6) represent unconventional substances, which do not fulfil the T-H criterion as they do not reproduce the ideal gas equation for diluted gases (V → ∞). This last result does not necessarily imply that those kinds of equations do not obey the first and second laws of thermodynamics; the only implication is that they do not reproduce the ideal gas in the diluted limit.

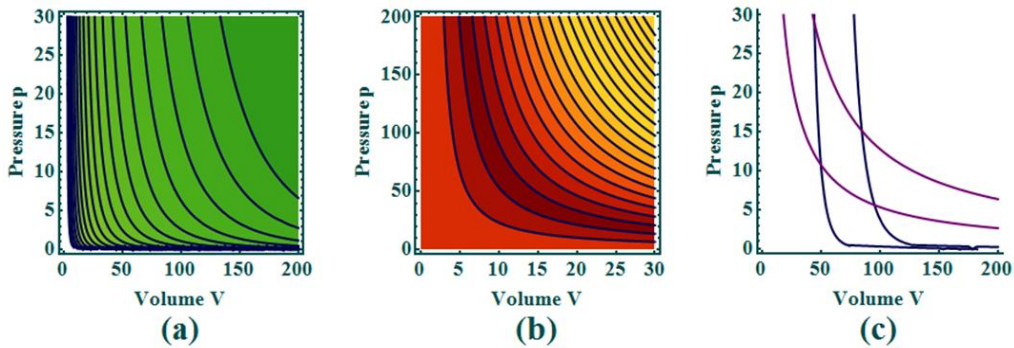

**Figure 3.** (a) Adiabatic processes, (b) isothermal processes and (c) a possible Carnot cycle for the state equation $p = \frac{aT}{V} + \frac{bT^2}{V^3}$.

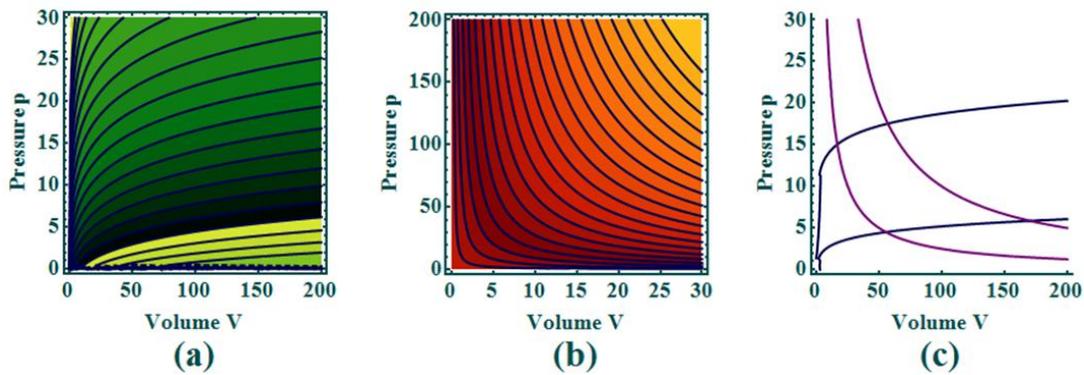

**Figure 4.** (a) Adiabatic processes, (b) isothermal processes and (c) a possible Carnot cycle for the state equation $p = \frac{aT^2}{V^2} + \frac{bT^4}{V^5}$.

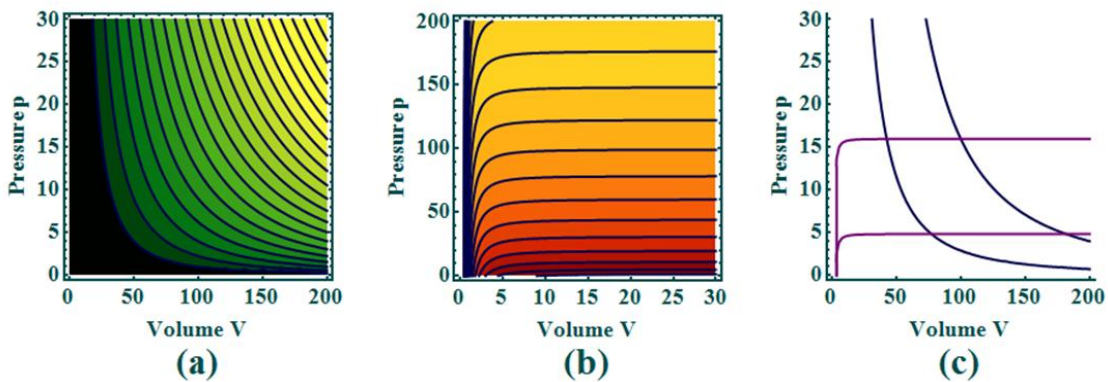

**Figure 5.** (a) Adiabatic processes, (b) isothermal processes and (c) a possible Carnot cycle for the state equation $p = aT^2 - \frac{b}{V^2}$.

We now analyse the unconventional particular behaviour (in addition to the comments stated above) of equation (2). If we use equations (9) and (19) for $U$ and $S$ and we substitute equation (2) into them, we obtain

$$U = -\frac{aV^2}{T} + \frac{3}{2}nRT = -\frac{Tp^2}{a} + \frac{3}{2}nRT, \qquad (26)$$

and

$$S = -\frac{aV^2}{2T^2} + \frac{3}{2}nR \ln T = -\frac{p^2}{2a} + \frac{3}{2}nR \ln\left(\frac{aV}{p}\right), \qquad (27)$$

where for simplicity the arbitrary temperature function is taken as $F(T) = \frac{3}{2}nRT$.

In figure (2c) we observe the two isotherms corresponding to $T_1$ and $T_2$ ($T_1 > T_2$). We also see the two adiabats joining the isotherms. Along the isotherm $1 \to 2$, we have,

$$\Delta U_{12} = -\frac{a}{T_1}(V_2^2 - V_1^2) < 0, \qquad (28)$$

and

$$W_{12} = -\int_1^2 p\,dV = -\frac{a}{2T_1}(V_2^2 - V_1^2) < 0. \qquad (29)$$

Thus, for $Q_1 = \Delta U_{12} - W_{12}$, we have,

$$Q_1 = -\frac{a}{2T_1}(V_2^2 - V_1^2) < 0. \qquad (30)$$

That is, along this process the system is rejecting heat. On the other hand, for the isotherm $3 \to 4$ at temperature $T_2 < T_1$ we have,

$$Q_2 = -\frac{a}{2T_2}(V_4^2 - V_3^2) > 0, \qquad (31)$$

that is, the system is absorbing heat. We can also obtain the total work as,

$$W_{TOT} = |Q_2| - |Q_1| = Q_2 + Q_1, \qquad (32)$$

from the first law with $\Delta U_{cycle} = 0$. By the substitution of equations (30) and (31) into (32), we have,

$$W_{TOT} = -\frac{a}{2T_2}(V_4^2 - V_3^2) - \frac{a}{2T_1}(V_2^2 - V_1^2). \qquad (33a)$$

This can be rewritten, using the state equation (2), as

$$W_{TOT} = -\frac{T_2}{2a}(p_4^2 - p_3^2) - \frac{T_1}{2a}(p_2^2 - p_1^2). \qquad (33b)$$

The equation for the adiabatic processes is (see appendix B),

$$\frac{e^{\frac{p^2}{3nRa}}}{T} = constant. \qquad (B6)$$

By applying the adiabatic equation to processes 2 → 3 and 4 → 1, we obtain,

$$p_4^2 - p_3^2 = p_1^2 - p_2^2 = \gamma, \tag{34}$$

where $\gamma$ is a positive constant. Equation (33b) can be rewritten as

$$W_{TOT} = -\frac{T_2}{2a}(p_4^2 - p_3^2) - \frac{T_1}{2a}(p_2^2 - p_1^2) = -\frac{\gamma}{2a}(T_2 - T_1) > 0. \tag{35}$$

Notice that the convention is that the total work $W$ is negative when the system makes work and positive when the work is made over the system. Therefore, this Carnot cycle works as a refrigerator, since heat goes from a lower temperature to a higher temperature. Obviously the inverse cycle is a direct Carnot engine with efficiency given by $\eta_C = 1 - T_2/T_1$ according to the procedure described in section 2. Thus, this hypothetical substance seemingly has an inverse behaviour with respect to conventional gases. In an isothermal expansion (equation (28)) an increment of pressure gives a negative change in the total internal energy $U$. Moreover, $U$ diverges to minus infinity when $T \to 0$, and by using equation (27) we see how in an isothermal process $S$ decreases when $p$ increases (more order or information when the volume is larger!). In fact, the state equation here discussed has a negative isothermal compressibility $\kappa_T = -1/ap$, but its coefficient of thermal expansion is $\beta = 1/T > 0$; that is, the same expression of $\beta$ for an ideal gas. Recently, in ref. [14] mechanical meta-materials with negative compressibility transitions have been discussed.

**4. The preserving areas property**

State equations obey Carnot theorem even when they are not supposed (according to some criteria) to do it. This interesting feature can be tracked all the way back through the obtention of the equations of state by asking about those identities that should be violated in order to not satisfy the theorem. With just first and second laws as valid facts, the only assumption made is the validity of the Maxwell relations (used in order to obtain Eq. (7)), which are mathematical constructions with some interesting outcomes. The formation of mappings that preserve areas is one of them. This property of area preservation (that will be shown below) has the first law of thermodynamics as a consequence for a cycle, i.e. $0 = Q + W$ (the change of the total internal energy is zero for a simple closed curve), which leads to Eq. (10). Maxwell relations also allow the calculation of the heat by giving the proper relationship between heat and entropy in terms of the pressure as can be seen in the obtention of Eq. (11). This indicates that Maxwell relations play a major role in the obtention of the Carnot efficiency treated with the equations of state. In this section the preserving area maps will be discussed resulting in extending the previously obtained calculations to another work conjugate variables. The conclusion is that within the Maxwell relations some particular cases of the first and second laws of thermodynamics are summarized. The explicit form of the Carnot efficiency lies on the change of coordinate systems which preserve area.

As was mentioned in the Introduction, every Carnot cycle depicted in the work plane ($Y - X$ plane) is mapped to a rectangle in its corresponding heat plane ($T - S$ plane). Obviously, the first law of thermodynamics demands that their corresponding cycle areas in both planes must be the same (for fixed $T_1$ and $T_2$). This result is equivalent to the fact that the Jacobian of $(p, V)$ with respect to $(T, S)$ is equal to one. Evidently, this property involving the Jacobian identities is also valid for any pair of conjugate variables defining a work plane, for example, $(H, M)$, $(F, l)$, $(\sigma, A)$, in the case of paramagnets, elastic wires, and surface films, respectively.

The explicit calculation of the Jacobian for every state equation is not always possible since in many cases the change of coordinate systems is mathematically impossible to determine. For example, for equation (3) it is not possible (because the solution for a polynomial of arbitrary degree is unknown), but in the particular cases of equations (2), (4), (5) and (6) the calculation of the Jacobian for the change of coordinate systems $(p, V)$ and $(T, S)$ gives indeed one.

It is well known that the Jacobian identities have the property of preserving region areas bounded by simple closed curves in both planes involved in the transformation of coordinate systems [15]. Another important property of Jacobian identities such as,

$$\frac{\partial(p, V)}{\partial(T, S)} = 1, \tag{36}$$

is the derivation of the Maxwell relations. For the case of equation (36) we have [11],

$$\frac{\partial(p, V)}{\partial(T, S)} = \frac{\partial(p, V)}{\partial(T, V)} \frac{\partial(T, V)}{\partial(T, S)} = \left(\frac{\partial p}{\partial T}\right)_V \left(\frac{\partial V}{\partial S}\right)_T = 1, \tag{37}$$

and then, we obtain the well-known Maxwell relation given by

$$\left(\frac{\partial S}{\partial V}\right)_T = \left(\frac{\partial p}{\partial T}\right)_V. \tag{38}$$

In equation (37), the auxiliary variables $T$ and $V$ were used. The Maxwell relations for simple hydrostatic thermodynamic systems are easily obtained from the exact differentials of thermodynamic potentials resulting from Legendre internal energy transformations of the system [16]. This result is very general. Let the function $M$ be defined in a subspace $V$ of $\mathbb{R}^2$ where $\vec{x} \in V$ is expressed as $\vec{x} = (w, z)$, then

$$M(\vec{x}) = M(w, z). \tag{39}$$

Its differential is

$$dM = m(w, z)dw + n(w, z)dz. \tag{40}$$

For simplicity we will just denote $m(w, z)$ and $n(w, z)$ as $m$ and $n$, then,

$$dM = mdw + ndz. \tag{41}$$

Let us take the Legendre transformation $K$ of $M$,

$$K = -wm + M, \tag{42}$$

in which we are exchanging the variable w for the variable m, related by,

$$m = \left(\frac{\partial M}{\partial w}\right)_z. \tag{43}$$

The differential of this new function is

$$dK = -wdm + ndz, \tag{44}$$

and because dK is exact, then by the Schwarz theorem,

$$\left(\frac{\partial w}{\partial z}\right)_m = -\left(\frac{\partial n}{\partial m}\right)_z. \tag{45}$$

Now let $\Omega$ be a mapping with $\Omega: (z, n) \mapsto (w, m)$ (in order to preserve areas the choice of coordinate systems is not arbitrary). It is easily shown that this mapping preserves areas (but changes orientations) given the fact that the Jacobian between these two coordinate systems is $-1$,

$$\frac{\partial(n, z)}{\partial(m, w)} = \frac{\partial(n, z)}{\partial(m, z)} \frac{\partial(m, z)}{\partial(m, w)} = \left(\frac{\partial n}{\partial m}\right)_z \left(\frac{\partial z}{\partial w}\right)_m, \tag{46}$$

and because of Equation (45)

$$\frac{\partial(n, z)}{\partial(m, w)} = -\left(\frac{\partial w}{\partial z}\right)_m \left(\frac{\partial z}{\partial w}\right)_m = -1. \tag{47}$$

This means that we have a preserving areas map between the planes $(n, z)$ and $(m, w)$.

For instance, in simple hydrostatic systems, usually one Maxwell relation is obtained from each thermodynamic potential $U$ (total internal energy), $F$ (Helmholtz free energy), $G$ (Gibbs free energy) and $H$ (enthalpy). These are [17]

$$\left(\frac{\partial T}{\partial V}\right)_S = -\left(\frac{\partial p}{\partial S}\right)_V, \tag{48a}$$

$$\left(\frac{\partial p}{\partial T}\right)_V = \left(\frac{\partial S}{\partial V}\right)_T, \tag{48b}$$

$$\left(\frac{\partial S}{\partial p}\right)_T = -\left(\frac{\partial V}{\partial T}\right)_p, \tag{48c}$$

$$\left(\frac{\partial T}{\partial p}\right)_S = \left(\frac{\partial V}{\partial S}\right)_p, \tag{48d}$$

respectively. The relation (48b) was obtained above (see equation (38)). For the other three Maxwell relations (37a,c,d) we can use exactly the same procedure shown in equation (37) but using different pairs of auxiliary variables in each case: for $U$, the pair $(p, S)$; for $G$, the pair $(V, T)$ and for $H$, the pair $(V, S)$.

A remarkable fact is that with the four variables involved in the right hand side of Gibbs equation $dU = TdS - pdV$, we can form six pairs. However, only the Jacobian between the $(p, V)$ and $(T, S)$ coordinate systems gives the identity which preserves areas. The other four pairs $(S, V)$, $(S, p)$, $(V, T)$, and $(T, p)$ have Jacobians that do not preserve areas, and they only work as auxiliary variables to calculate the four Maxwell relations. This result can be generalized for other thermodynamic systems with more than two degrees of freedom. For example, for the paramagnetic ideal gas, we have $dU = TdS - pdV + HdM$, and with six variables (on the right hand side) we can construct fifteen pairs, but with three pairs involved in preserving areas maps (two work planes, the $(H, M)$ and the $(p, V)$ planes, and the heat plane). In this way we can obtain twelve Maxwell relations, as is known for this system [3]. The Jacobian identities can be obtained by considering equation (47). That is, three preserving area maps and then twelve Maxwell relations for the paramagnetic ideal gas. When $V$ is constant, we can see that,

$$\frac{\partial(H,M)}{\partial(T,S)} = -1, \tag{49}$$

when $M$ is constant we have that,

$$\frac{\partial(p,V)}{\partial(T,S)} = 1, \tag{50}$$

and when $S$ is constant,

$$\frac{\partial(p,V)}{\partial(H,M)} = 1. \tag{51}$$

We believe that this rule can be employed for many other cases.

5. Conclusions
The celebrated Carnot theorem has been widely recognized in scientific literature for its deep meaning and consequences. It is a cornerstone in the thermodynamics building. Recently, Penrose reminded us of the universal magnificence of the second law. However, it is not easy to find work on this topic. In the present article we revisit this subject within the context of unconventional state equations for gases and by means of a geometrical property of simple closed cycles in the heat and work planes that is related with energy conservation.

For the analysis of equations of state we have proposed a somewhat general state equation for gases which can reproduce the virial and some unconventional state equations. The proposed function fulfils the Carnot efficiency, and therefore any particular equation deriving from it also verifies this, including the unconventional state equations and even some equations that are not consistent with thermodynamics according to the criterion for gases proposed by Tykodi and Hummel. The unconventional equations of state present uncommon behaviours, for example: the pressure in the very dilute limit is not zero, thus when the particles of the gas are far apart from one another, the interactions between them may not be zero; but instead, the pressure is proportional to some power of the temperature or a positive power of the volume; also in an isothermal expansion the heat could go from the cold to the hot reservoir and the entropy could diminish with an expanding volume. Surprisingly, these substances also fulfil the Carnot theorem.

At the same time, we study thermodynamic cycles under the perspective of a geometrical property within the context of areas preservation and the Jacobian identities. For a given thermodynamic system characterized by its Gibbs equation there are some pairs of thermodynamic variables, which have to be conjugate variables, whose Jacobian is equal to one; this means that there is a preserving area property between the work planes and the heat plane for a simple closed cycle (without using the first law of thermodynamics). This area property is summarized in the Maxwell relations.

As usual, the Maxwell relations are constructed through the Legendre transformation of the total internal energy, and it is shown that this transformations (with a suitable choice of coordinate systems) will lead to the preserving areas transformation. The choice of coordinates is not arbitrary, since the energy conservation is established only between the planes of conjugate variables. All this leads to a rule which consists in determining the number of Maxwell relations and the relations themselves.

The meeting point between the preserving areas property (meaning conservation of energy but not derived from first law) and the Carnot theorem are the Maxwell relations, which establish the proper connection between the heat plane and the working plane. Even in the case when the Carnot efficiency is obtained in the heat plane, one must invoke the fact that there is an energy conservation between the heat and the work planes, then proving the relevance of the Maxwell relations in the thermodynamics construction.

**Appendix A**

Determining the virial coefficients of gas state equation is a very difficult task if not impossible [12]. For known cases those coefficients can be expressed in a power series of the temperature but not necessarily with integer exponents. Here we will explicitly show that in the case that the virial coefficients could be written in one of this series, the virial expansion can be expressed by Eq. (3) with some issues to take into account. For example, consider the virial expansion for the Lennard-Jones interaction potential, which according to [13] has the following second and third virial coefficients,

$$B_2 = \sum_{k=0}^{\infty} d_{2k} T^{-(2k+1)/4},$$

$$B_3 = \sum_{k=0}^{\infty} d_{3k} T^{-(k+1)/2}.$$

(A1)

Let us suppose that the rest of the coefficients have a similar representation. Since $T > 0$, and $k \geq 0$, each term can be represented in a Taylor series expansion around some point $T'_0$, except for $T'_0 = 0$, then

$$T^{-(2k+1)/4} = \sum_{i=0}^{\infty} \frac{d^i T^{-(2k+1)/4}}{i! dT^i}\bigg|_{T'_0} (T - T'_0)^i = \sum_{i=0}^{\infty} g_{ki}(T - T'_0)^i$$

$$= T'_0{}^{-(2k+1)/4} - \frac{(2k+1)}{4} T'_0{}^{-(2k+5)/4}(T - T'_0) + \frac{(2k+1)(2k+5)}{32} T'_0{}^{-(2k+9)/4}(T - T'_0)^2 + \cdots$$

(A2)

and

$$T^{-(k+1)/2} = \sum_{l=0}^{\infty} \frac{1}{l!} \frac{d^l}{dT^l}\left(T^{-(k+1)/2}\right)\bigg|_{T'_0} (T - T'_0)^l = \sum_{l=0}^{\infty} g'_{kl}(T - T'_0)^l$$

$$= T'_0{}^{-(k+1)/2} - \frac{(k+1)}{2} T'_0{}^{-(k+3)/2}(T - T'_0) + \frac{(k+1)(k+3)}{8} T'_0{}^{-(k+5)/2}(T - T'_0)^2 + \cdots$$

(A3)

Then Eq. (1) can be rearranged as

$$p = RT\left(\frac{1}{v} + \frac{1}{v^2} B_2 + \frac{1}{v^3} B_3 + \cdots\right)$$

(1)

$$= RT\left(\frac{1}{v} + \frac{1}{v^2} \sum_{k=0}^{\infty} d_{2k} T^{-(2k+1)/4} + \frac{1}{v^3} \sum_{k=0}^{\infty} d_{3k} T^{-(k+1)/2} + \cdots\right)$$

$$= RT\left(\frac{1}{v} + \frac{1}{v^2} \sum_{k=0}^{\infty} \sum_{i=0}^{\infty} d_{2k} g_{ki}(T - T'_0)^i + \frac{1}{v^3} \sum_{k=0}^{\infty} \sum_{i=0}^{\infty} d_{3k} g_{ki}(T - T'_0)^i + \cdots\right),$$

$$= RT \sum_{j=0}^{\infty} \sum_{k=0}^{\infty} \sum_{i=0}^{\infty} d_{jk} g_{ki}(T - T'_0)^i v^{-j},$$

(A4)

if we define

$$a_{ij} = n^{-j} \sum_{k=0}^{\infty} d_{jk} g_{ki} \tag{A5}$$

where $n$ is the molar number. Then Eq. (A4) and (A5) lead to,

$$p = RT \sum_{j=0}^{\infty} \sum_{i=0}^{\infty} a_{ij}(T - T'_0)^i V^{-j} \tag{A6}$$

that would be the same than Eq. (3) when the sum is infinite and $a_{ij} = 0$ for $i < 0$ and $j > 0$ and $V_0 = 0$, except that in Eq. (A6) we have a $T$ factor multiplying all the expression. Nonetheless, expanding each $(T - T'_0)^i$ in Eq. (A6) according to the binomial theorem,

$$(T - T'_0)^i = \sum_{l=0}^{i} \frac{i!}{l!(i-l)!} T^l(-T'_0)^{i-l} = \sum_{l=0}^{i} \binom{i}{l} T^l(-T'_0)^{i-l},$$

allow us to write Eq. (A6) as

$$p = \sum_{j=0}^{\infty} \sum_{i=0}^{\infty} \sum_{l=0}^{i} \binom{i}{l} RT^{l+1}(-T'_0)^{i-l} a_{ij} V^{-j} \tag{A7}$$

by defining a new constant $c_i$ given by

$$c_i = \sum_{l=0}^{i} \binom{i}{l} R(-T'_0)^{i-l},$$

Eq. (A7) can be rewritten in the following form,

$$p = \sum_{j=0}^{\infty} \sum_{i=0}^{\infty} \sum_{l=0}^{i} a_{ij} c_i T^{l+1} V^{-j}. \tag{A8}$$

If we express

$$\sum_{i=0}^{\infty} \sum_{l=0}^{i} a_{ij} c_i T^{l+1} = \sum_{k=1}^{\infty} a'_{kj} T^k, \tag{A9}$$

then Eq. (A9) is

$$p = \sum_{j=0}^{\infty} \sum_{k=1}^{\infty} a'_{kj} T^k V^{-j}, \tag{A10}$$

which now is the same than Eq. (3) when the sum is infinite and $a_{ij} = 0$ for $i \leq 0$, $j > 0$ and $V_0 = T_0 = 0$. Taking care not to confuse the $T_0 = 0$ in Eq. (3) from the $T'_0 \neq 0$ which is the reference point in the Taylor expansion used in Eqs. (A2) and (A3), in fact, this representation (Eq. (A10)) is only valid for $T \neq 0$, because in $T = 0$ the original function is undetermined.

**Appendix B**

The state equation is the following:

$$p = \frac{aV}{T}. \tag{B1}$$

The inexact differential of heat is

$$dQ = \left(\frac{\partial U}{\partial T}\right)_V dT + \left[\left(\frac{\partial U}{\partial V}\right)_T + p\right] dV = 0, \tag{B2}$$

which is zero for an adiabatic process. From equations (7) we know that $\left(\frac{\partial U}{\partial V}\right)_T = T\left(\frac{\partial p}{\partial T}\right)_V - p$ and from equation (26), $\left(\frac{\partial U}{\partial T}\right)_V = \frac{aV^2}{T^2} + \frac{3}{2}nR$. Then

$$0 = \left(\frac{aV^2}{T^2} + \frac{3}{2}nR\right) dT + T\left(\frac{\partial p}{\partial T}\right)_V dV = \left(\frac{aV^2}{T^2} + \frac{3}{2}nR\right) dT - \frac{aV}{T} dV, \tag{B3}$$

that can be rewritten as

$$0 = \left(\frac{p^2}{a} + \frac{3}{2}nR\right) dT - p\left(\frac{pdT + Tdp}{a}\right) = \frac{3}{2}nRdT - \frac{Tp}{a} dp, \tag{B4}$$

therefore,

$$3nRa\frac{dT}{T} = 2pdp, \tag{B5}$$

arriving finally at the trajectory for an adiabatic process:

$$\frac{e^{\frac{p^2}{3nRa}}}{T} = constant. \tag{B6}$$

**Acknowledgements**
We thank R. Hernández-Pérez and T. E. Govindan for reading the manuscript. We also thank the referee for his valuable suggestions. This work was partially supported by CONACYT, PIFI-SIP, COFAA and EDI-IPN Mexico.